\documentclass[superscriptaddress,onecolumn,notitlepage,howpacs,aps]{revtex4-1}

\usepackage{graphicx}
\usepackage{dcolumn}
\usepackage{bm}
\usepackage{hyperref}
\usepackage{url}
\usepackage{color}
\usepackage{amsmath}

\definecolor{reddeep}{rgb}{0.995,0.000,0.200}

\begin{document}
\title{Phase transition of chemically doped uniaxial relaxor ferroelectric}

\author{S. Chillal}
\affiliation{Neutron Scattering and Magnetism, Laboratory for Solid State Physics, ETH Z\"urich, Z\"urich, Switzerland}

\author{D. Koulialias}
\affiliation{Neutron Scattering and Magnetism, Laboratory for Solid State Physics, ETH Z\"urich, Z\"urich, Switzerland}

\author{S.N. Gvasaliya}
\email[]{sgvasali@phys.ethz.ch}
\affiliation{Neutron Scattering and Magnetism, Laboratory for Solid State Physics, ETH Z\"urich, Z\"urich, Switzerland}

\author{R.A. Cowley}~\thanks{Deceased.}
\affiliation{Clarendon Laboratory, Department of Physics, Oxford University, Parks Road, Oxford, OX1 3PU, UK}

\author{L.I. Ivleva}
\affiliation{Prokhorov General Physics Institute, 119991, Moscow, Russia}

\author{S.G. Lushnikov}
\affiliation{Ioffe Physico-Technical Institute RAS, 194021, St. Petersburg, Russia}

\author{A. Zheludev}
\affiliation{Neutron Scattering and Magnetism, Laboratory for Solid State Physics, ETH Z\"urich, Z\"urich, Switzerland}

\begin{abstract}
We report a neutron scattering study of the ferroelectric phase transition in
 Sr$_{0.585}$Ce$_{0.025}$Ba$_{0.39}$Nb$_2$O$_6$ (SBN-61:Ce). We find no evidence for a soft transverse 
optic phonon. We do, however, observe anisotropic diffuse scattering. This scattering has inelastic and 
elastic contributions. In the paraelectric phase the susceptibility associated with the elastic diffuse 
scattering from SBN-61:Ce increases on approaching the transition temperature. 
In the ferroelectric phase the lineshape of the elastic scattering is consistent with the form expected for 
the ferroelectric domain walls. 
In contrast to the macroscopic observations, the scattering properties of Ce-doped crystal 
do not exhibit important changes with respect to those of pure Sr$_{0.61}$Ba$_{0.39}$Nb$_2$O$_6$. 

\end{abstract}
\pacs{{77.80.B} {Phase transitions}; {61.05.fg} {Neutron scattering}; {64.60.Cn,81.30.Hd} {Order-disorder transitions}; 
}

\date{\today}

\maketitle

\nopagebreak[1]

\section{Introduction}

Relaxor ferroelectrics, shortly relaxors, are disordered crystals with outstanding dielectric, piezoelectric and 
electro-optic properties~\cite{blinc2011advanced}. 
It is generally believed that disorder is a key ingredient in the physics of relaxors, but a consistent 
model of their behavior has still not been
developed~\cite{PhysRevLett_111_097601_levanyuk,PhysRevLett_110_147602_rappe,PhysRevLett.111.227601,PhysRevB.88.134106,
advphys_60_229,Phelan04022014,ISI:000335221900002,PhysRevB.91.144105}. Even the basic criteria for a disordered ferroelectric 
to become a relaxor have not yet been established. 

Strontium Barium Niobate, Sr$_x$Ba$_{1-x}$Nb$_2$O$_6$ (SBN), is a group of disordered uniaxial ferroelectrics 
with unfilled tetragonal tungsten bronze structure~\cite{sbn1668176}. The phase transition of SBN is strongly 
affected by the relative amount of Sr/Ba ions or via doping with rare-earth ions. The effects of the latter 
chemical modification are most often studied for Sr$_{0.61}$Ba$_{0.39}$Nb$_2$O$_6$ (SBN-61) as large and 
high-quality single crystals of this composition are available.    
The observation for SBN-61 is that even a minuscule concentration of Ce$^{3+}$ ions affects strongly the ferroelectric 
phase transition~\cite{ISI:000086918400004}. More specifically, the anomaly in dielectric permittivity $\varepsilon$ 
broadens and its frequency dispersion becomes much more pronounced in SBN-61:Ce as compared to the pure crystal. The peak 
in $\varepsilon$ shifts towards lower temperatures upon increasing the doping level. 
The ferroelectric polarization is history-dependent near the transition 
temperature~\cite{PhysRevLett.92.065701,PhysRevLett.97.065702}. There is a strong change in polarization aging 
under light illumination in SBN-61:Ce~\cite{ref_2002}. Thus, the macroscopic properties of 
doped SBN-61 crystals have much in common with relaxors. 

The phase transition of pure SBN-61 has been recently studied by neutron 
scattering~\cite{pss_55_334,sbn61,PhysRevLett.113.167601}. On approaching from above the transition temperature, 
T$_C$, a slow dynamic diffuse scattering (DS) emerges in the neutron spectra ~\cite{sbn61,PhysRevLett.113.167601}. 
It appears to account for divergent dielectric permittivity~\cite{sbn61} 
and has a characteristic energy scale in the $\mu$eV range ~\cite{PhysRevLett.113.167601}. Well below 
T$_C$ the DS appears in neutron spectra as elastic component and 
can be fully ascribed to ferroelectric domain walls (DW)~\cite{sbn61}. Thus, despite inherent chemical disorder 
the scattering from SBN-61 is similar to the one from classic order-disorder ferroelectric. 
It is therefore tempting to establish if doping, which considerably modifies the macroscopic  
properties of SBN-61, leads to substantial changes in the microscopic properties as seen by neutrons.
We report here a neutron scattering study of the ferroelectric phase transition in 
Ce-doped Sr$_{0.61}$Ba$_{0.39}$Nb$_2$O$_6$ single crystal. In stark contrast with modifications of the 
macroscopic properties induced by doping, we do not find important changes in the scattering properties 
with respect to the parent pure material.  

\section{Sample \& the Experimental Details}

\noindent Sr$_{0.61}$Ba$_{0.39}$Nb$_2$O$_6$ has a tetragonal structure. Above T$_\mathrm{c}$ the space group of 
SBN is $4/mmm$, while below the ferroelectric phase transition temperature the space group is $4mm$. The 
unit cell contains 5 formula units, so that five Sr/Ba ions are distributed over six available positions. 
These two ions have considerably different ionic radii 
(r$\rm_{Sr^{2+}}$ = 1.12 \AA~and r$\rm_{Ba^{2+}}$ = 1.34 \AA )~\cite{Shannon:a12967}. 
The structure of SBN is a three dimensional network of distorted Nb/O octahedra connected together so that 
there are pentagonal, square and triangular tunnels. The smallest triangular tunnels are not occupied. Only 
strontium ions reside at the square tunnels. Finally, strontium and barium ions 
are distributed over the largest pentagonal channels. As the Sr$^{2+}$ and Ba$^{2+}$ ions have formally the 
same charge there are no strong electrostatic forces related with the distribution of the Sr and Ba ions. 
The possible charge irregularity is then due to presence of unfilled positions in the triangular channels. 

An obvious charge disorder is introduced in SBN upon doping it with trivalent cerium. 
Ce$^{3+}$ ions occupy Sr$^{2+}$ sites in the lattice~\cite{ISI:000077482700013}, although there are evidences for their 
slight shifts with respect to the nominal positions of strontium~\cite{jpcmce}. The Ce$^{3+}$ ions are likely to cause 
substantial electrostatic forces related to simultaneous charge and chemical randomness. Such a modification brings 
SBN closer to the Pb-based cubic relaxors, such as PbMg$_{1/3}$Nb$_{2/3}$O$_3$ (PMN).  
\begin{figure}
\includegraphics[width=0.5\columnwidth]{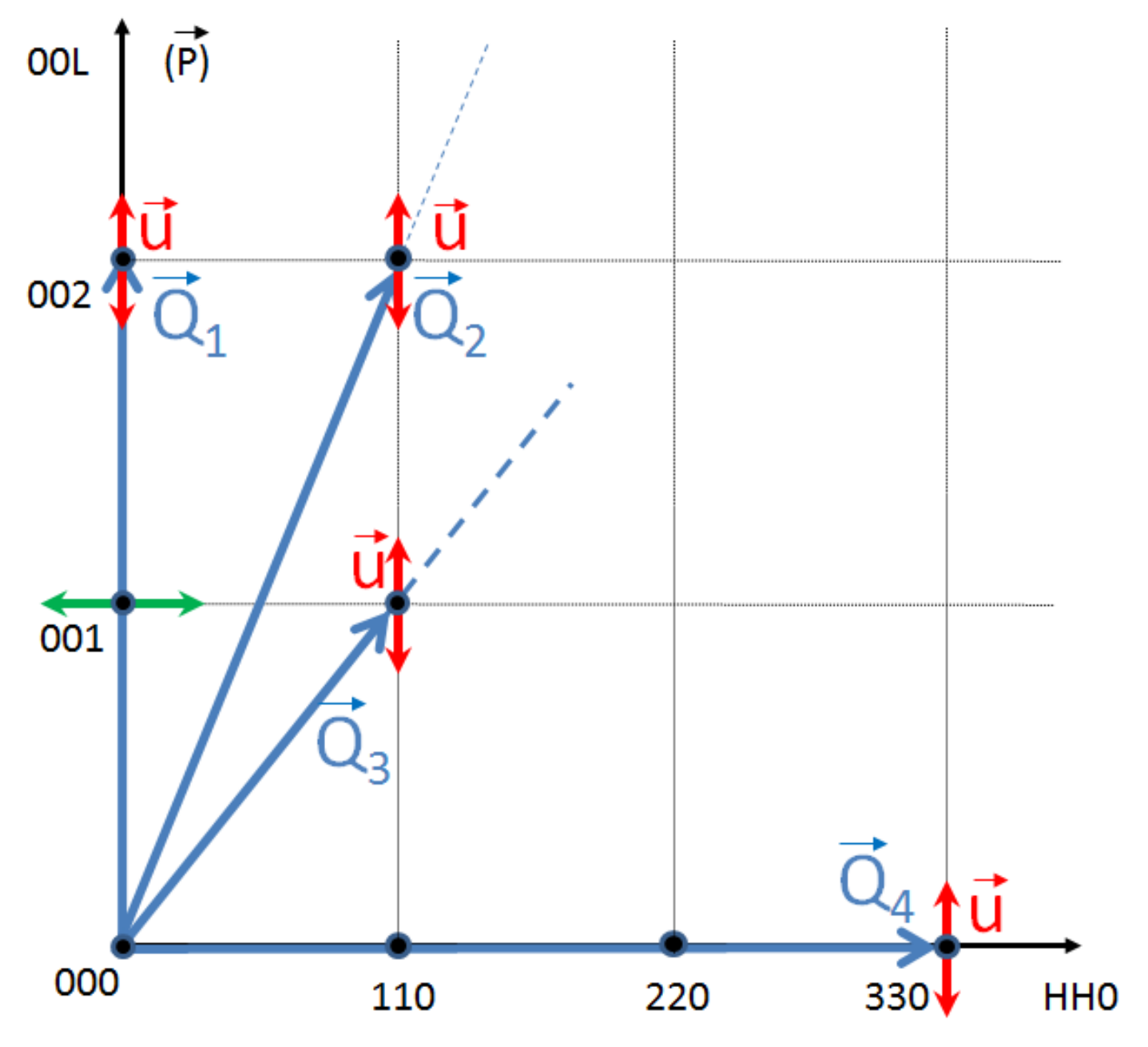}
\caption{A sketch of the scattering plane used in our experiments. The scattering vectors $\mathbf{Q}_i$ denote the Bragg 
         positions where the distributions of neutron intensities were collected. The polarization vector $\mathbf{u}$ 
         of the ionic displacements related to the diffuse scattering is shown in red. 
         The green arrow denotes the direction of (transverse) scans A direction of spontaneous 
         polarization (a {\it direct space vector}) is schematically shown as $\pm\mathbf{P}$} 
\label{scatplane}
  \end{figure}

We performed a neutron scattering study of Sr$_{0.61}$Ba$_{0.39}$Nb$_2$O$_6$ doped by 2.5 atomic \%Ce, 
Sr$_{0.585}$Ce$_{0.025}$Ba$_{0.39}$Nb$_2$O$_6$ (SBN-61:Ce). The crystal was grown by using modified Stepanov  
process~\cite{INSPEC:12551392}. A congruently melting composition of Sr$_{0.61}$Ba$_{0.39}$Nb$_2$O$_6$ was doped by 
0.1 wt.\% CeO$_2$. High quality single crystals grow even at this rather large concentration of Ce ions~\cite{ceuni}. 
The ferroelectric phase transition temperature for this crystal was estimated from dielectric measurements to be 
T$_\mathrm{c}\sim$~348~K. The precise determination of T$_\mathrm{c}$ for doped crystals is problematic as the increase 
in polarization or in linear optical birefringence turns out to be less sharp in 
temperature~\cite{AIP2000,PhysRevLett.97.065702}. Nevertheless, the T$_\mathrm{c}\sim$~348~K is also 
inferred from birefringence results for a crystal with amount of Ce ions nearly identical to our 
sample~\cite{AIP2000}. 

We used a single crystal of SBN-61:Ce with dimensions of $3\times3\times1.3$ cm$^3$. The mosaic spread was 
within the resolution of the spectrometer. The sample was aligned in the $<$0,0,1$>$/$<$1,1,0$>$ scattering plane. 
This experimental geometry is sketched in Fig.~\ref{scatplane}. 
The lattice parameters of SBN-61:Ce at T=300 K are $a = b= 12.43$ \AA, and $c = 3.93$ \AA. The crystal was mounted 
in a displex refrigerator that enabled the temperature to be controlled between 20~K and 500~K. 

The experiments were conducted with the cold neutron 3-axis spectrometer TASP~\cite{Semadeni2001152}, 
situated at the end of a curved guide at the SINQ facility (PSI, Switzerland). The energy of the scattered neutrons 
was kept fixed to 8.04~meV and a PG filter was installed in front of the analyzer. Most of the data was collected 
with the collimation in the horizontal plane from reactor to detector as open-80$'$-sample-80$'$-80$'$, giving an 
energy resolution of 0.40 meV. Some data was taken with a tighter 
collimation: open-20$'$-sample-20$'$-20$'$, improving the resolution to 0.2 meV.

\section{Results}

\subsection{Low-energy excitations across T$_\mathrm{c}$}
\label{lowe}
 
\begin{figure}
\includegraphics[width=0.5 \columnwidth]{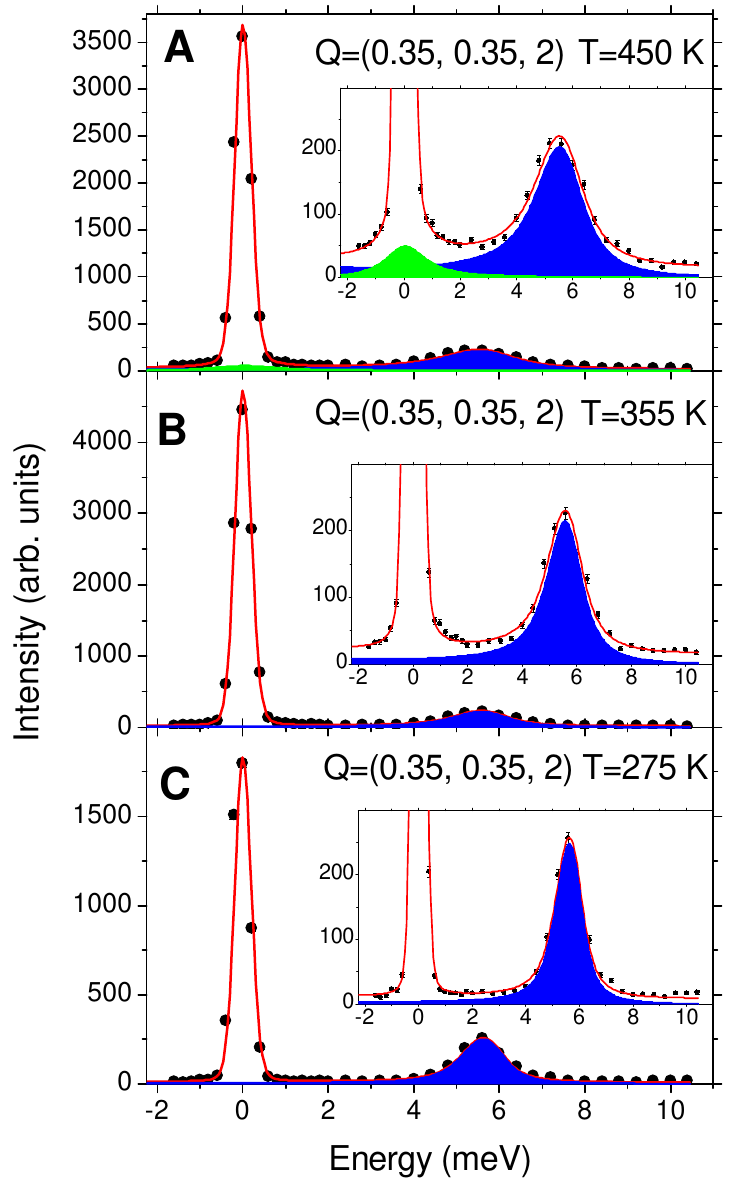}
\caption{Typical neutron spectra from SBN-61:Ce taken at the temperatures above, approximately at T$_\mathrm{c}$ and below the 
         phase transition at wavevector $\mathbf{Q}$=(0.35,~035,~2). The data presented 
         in Fig.~\ref{shapeinel} is obtained with open-80$'$-80$'$-80$'$ collimation. The solid circles show the data, the 
         red lines give the results from the best-fits, blue and green areas emphasize contributions of the TA phonon and 
         the dynamic diffuse scattering. Sharp in energy and 
         very intense contribution from the EDS at elastic position is self-evident.
         The data was taken with collimation as open-80$'$-80$'$-80$'$. } 
         \label{shapeinel}
    \end{figure}

Fig.~\ref{shapeinel} shows the neutron spectra for a wave-vector transfer $\mathbf{Q}=(0.35, 0.35, 2)$ at temperatures 
above, near, and below the phase transition T$_\mathrm{c}\sim$348~K. This data shows two peaks. One of them is centered at 
zero energy transfer, while the other one is an 
inelastic peak. The inelastic peak shifts towards higher energies and broadens at higher wavevector transfers $\mathbf{Q}=(q, q, 2)$. 
Qualitatively, the most pronounced changes in these spectra are the significant variations in elastic intensity 
and a reduction in width of the inelastic peak upon decrease of temperature. No intensity that could be 
associated with a low-energy transverse 
optic phonon is observed, despite the frequency range of our scans covers positions of several optic phonons detected in SBN by 
infrared reflectivity~\cite{0953-8984-17-4-008} and Raman spectroscopy~\cite{doi:10.1080,1.358943}. 
\begin{figure*}
\includegraphics[width=1.00\columnwidth]{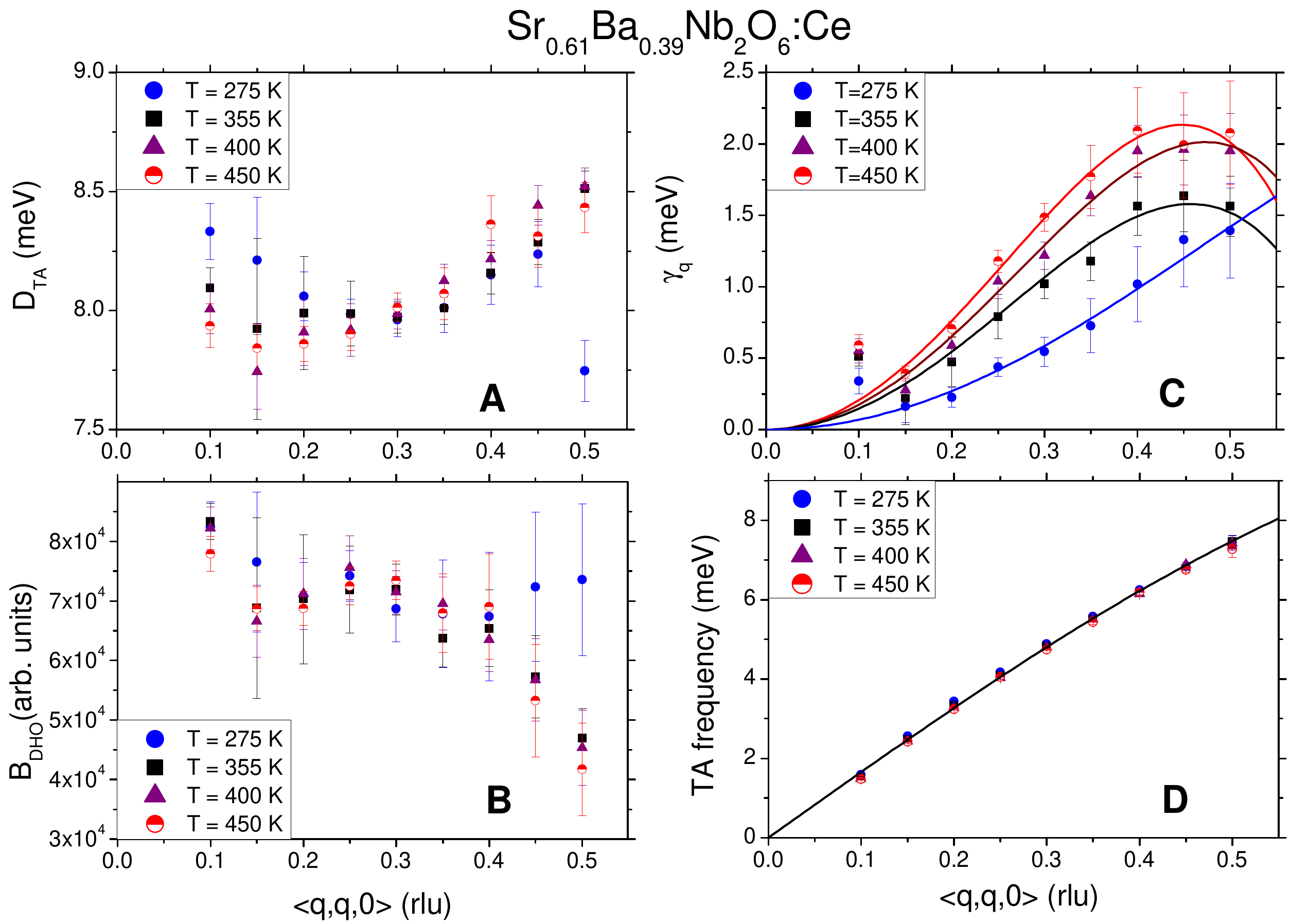}
\caption{(A-C) Wavevector dependence of the parameters of DHO inferred from the data taken in the vicinity of the 
              (0,0,2) Bragg peak at several temperatures. (D) A dispersion curve of 
              the TA phonon obtained from simple fits by a Lorentzian.  
              The apparent difference in the quality of the results shown in Figs.~\ref{parinel}A and~\ref{parinel}D is due 
              to a broader range of parameters 
              spanned by the dispersion curve plotted in Fig.~\ref{parinel}D. The error bars for 
              the parameters shown in Figs.~\ref{parinel}A and~\ref{parinel}D are comparable and amount to $\sim$2.5\% 
              on average. The lines in the panel (C) are the fits to the phenomenological expression  
              $\gamma_q(T)$=$\gamma_2(T)\cdot q^2 + \gamma_4(T)\cdot q^4$ and serve as a guide to the eye. rlu stands for 
              reciprocal lattice unit.} 
  \label{parinel} 
\end{figure*}

For quantitative analysis of the neutron data we develop an appropriate model. First attempts were made to describe the spectra 
by a sum of a Gaussian peak at 
zero energy transfer and a damped harmonic oscillator (DHO). This model failed to describe the experimental data satisfactorily. 
The deficiency of this scattering function was especially pronounced at low energy transfers, at the tails of the peak 
centered at zero energy. These tails suggested 
importance of an additional, resolved in energy, contribution. The model was therefore 
extended by including a Lorentzian peak centered on zero energy transfer:
\begin{eqnarray}
\label{sf_phonon}
S(\mathbf{Q},\omega) & = & A_{EDS}(\mathbf{Q})\delta(\omega)+\frac{1}{\pi}(n(\omega)+1)\cdot\nonumber \\
& \cdot& (B(\mathbf{Q})_{DHO} 
\cdot \chi_{DHO}+C_{IDS}(\mathbf{Q})\cdot\chi_{IDS})'' 
\end{eqnarray} 
\noindent In Eq.~\ref{sf_phonon} the first term describes the elastic, limited in energy by experimental resolution, 
diffuse scattering (EDS). 
The second term accounts for phonon scattering. The last term approximates the energy-resolved  
diffuse scattering (IDS). The components with a finite-energy width have to be combined with the detailed balance 
factor $(n(\omega)+1)=(1-exp[-\hbar\omega/(k_bT)])^{-1}$. $\chi_{DHO}=(\omega_{TA}^2-i\gamma_q\omega-\omega^2)^{-1}$ 
and $\chi_{IDS}=(1-i\omega/\Gamma_{IDS})^{-1}$ are the wavevector-dependent susceptibilities of the phonon 
and of the IDS. The dispersion of the TA phonon was approximated as $\omega_{TA}=D_{TA}\cdot\sin(0.5 \pi q)$, 
where $q=\sqrt{q_x^2+q_y^2}$. The stiffness $D_{TA}$, damping $\gamma_q$, $\Gamma_{IDS}$, and the scale factors 
$A_{EDS}$, $B(\mathbf{Q})_{DHO}$, and $C_{IDS}(\mathbf{Q})$ were allowed to vary as a function of wavevector without restrictions.  
The scattering function Eq.~\ref{sf_phonon} was convoluted with the resolution function by using 
the ResLib4.2 library~\cite{reslib}. A constant background was finally added to the intensity. 
The fitted results gave a good description of the data as shown in Fig.~\ref{shapeinel}.

\noindent We now describe the temperature evolution of the components in the low-energy neutron spectra from SBN-61:Ce. 
The parameters of the DHO function used to approximate the TA phonon are shown in Fig.~\ref{parinel} as a function of wavevector 
for several temperatures. As is clearly seen  
in Fig.~\ref{parinel}A the stiffness of the TA phonon across the T$_\mathrm{c}$ does not change within the errors. No scattering 
that could be associated with an optic phonon was detected. This suggests that 
there is no softening of a low-energy optic branch. In other words, we do not find any evidence for noticeable decrease 
in the frequencies of the phonons upon approaching the transition temperature. It might happen, however, 
that such a softening of an optic phonon is not directly observed in neutron 
spectra due to unfavorable structure factor around a particular reciprocal lattice point. This hidden softening could still 
significantly distort the 
lineshape of the TA phonon and thus change its structure factor and/or damping~\cite{PhysRevLett36806,advphys_1980_1}. As 
follows from the temperature evolution of these parameters shown in Figs.~\ref{parinel}(B,C) this is not the case 
in SBN-61:Ce. The (squared) structure factor of the TA phonon is fairly 
temperature-independent. The damping of the TA phonon simply decreases towards lower temperatures; 
this behavior is just as expected for a weakly 
anharmonic system. The energy resolved diffuse scattering is weak as can be seen from Fig.~\ref{shapeinel}. As a function 
of wavevector the intensity and the width $\Gamma_{IDS}$ 
of IDS do not show any regular dependence. The intensity of the IDS just gradually diminishes at lower temperatures.

Altogether, our results show that 
energy-resolved components in the neutron spectra of SBN-61:Ce do not play an important r$\rm \hat{o} $le in the ferroelectric phase 
transition of this crystal. As is obvious from Fig.~\ref{shapeinel} the most noticeable changes in the spectra are in 
the elastic, limited in energy by the 
spectrometer resolution, component. At $\mathbf{Q}=(0.35, 0.35, 2)$ the peak intensity of the EDS varies by a factor 3 
in the considered temperature range. 
One should therefore conclude that the critical dynamics in SBN-61:Ce is concentrated at low energies.

\subsection{Shape of neutron diffuse scattering from SBN-61:Ce}
\label{shape}

%
\begin{figure*}
\includegraphics[width=1.0 \columnwidth]{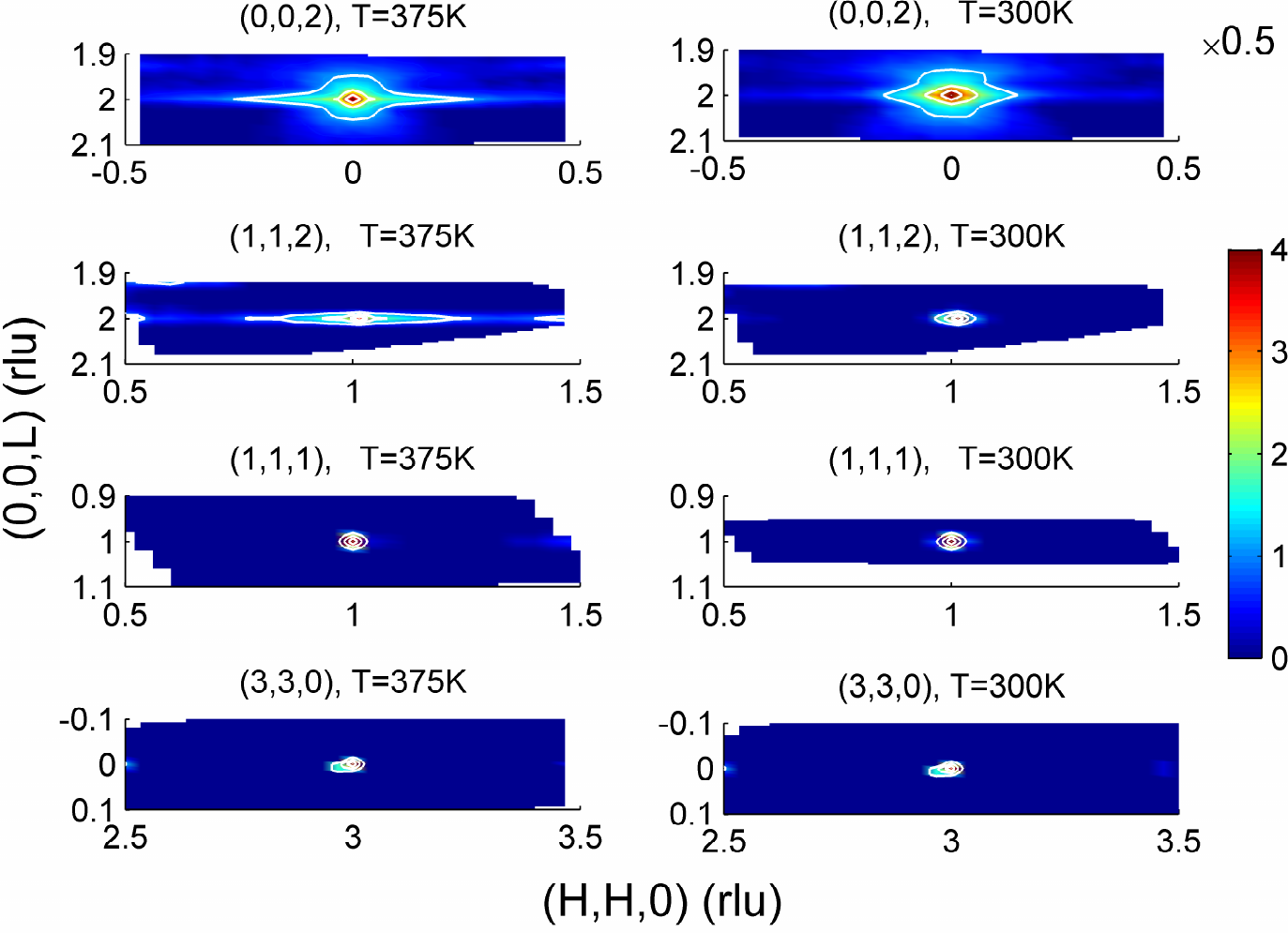}
\caption{False-color plots of the distribution of neutron elastic intensity in the paraelectric 
        (left column) and in the ferroelectric (right column) 
         phases. The maps in the paraelectric phase were collected at T~=~375~K. The measurements in the 
         ferroelectric phase were performed at T~=~300~K. The intensities are presented on a  
         logarithmic scale. The overall intensities around the (0~,0~,2) Bragg position are considerably higher 
         than that around other reciprocal lattice points. To maintain a single color-bar for the entire figure 
         the range of the color space for the 
         data in the upper raw is reduced by a factor of two. 
         The data was taken with collimation as open-80$'$-80$'$-80$'$.} 
  \label{maps}
\end{figure*}
\noindent SBN-61:Ce is considerably disordered crystal. In such a case there might be several equally important sources of EDS. 
To get a hint on the origin of the EDS that showed strong temperature dependence we analyze the distributions of 
the EDS from SBN-61:Ce near 
several Bragg peaks above and below the ferroelectric phase transition. Corresponding false color plots for the data 
taken at T~=~375~K and at T~=~300~K 
are shown in Fig.~\ref{maps}. Around the (0,0,2) and the 
(1,1,2) Bragg positions at T~=~375~K the EDS in the form of lines extended along the $<1,1,0>$ direction is observed. 
However, no similar EDS is detected around the (1,~1,~1) and (3,~3,~0) Bragg peaks. 
Such shape of the EDS is generally associated with 
pronounced quasi-1D ionic correlations along the ferroelectric $c$-axis. Indeed, in the latter case a diffuse scattering 
would form the planes (sheets) in the reciprocal space that are 
orthogonal to the direction with pronounced correlations. In the geometry of our experiments these sheets appear 
as sharp lines perpendicular to the $c$-axis. 
Similar distribution of the EDS is actually observed in our data taken in the vicinity of the (0,~0,~2) and (1,~1,~2) positions. 
Also, the absence of the EDS around the (3,3,0) position is a natural consequence of the displacive 
nature of the quasi-1D correlations. Indeed, 
for {\it any} scattering associated with ionic displacements the intensity should contain the 
polarization factor~\cite{krivoglaz1996diffuse} 
of the form $\lvert\mathbf{Q\cdot u}\rvert^2$, where $\mathbf{Q}$ is the total wavevector transfer and $\mathbf{u}$ 
is the pattern of the 
ionic displacements. Clearly, for the displacements occurring along the $c$-axis, this factor entirely suppresses 
the intensity near the positions of the 
(h,h,0)-type. In addition, the significant difference in the EDS near the (0,~0,~2) and the (1,~1,~2)\& (1,~1,~1) 
positions suggests a complex pattern 
of the ionic displacements along the $c$- axis. Let us compare the EDS  intensities taken at T=~375~K along 
the $<1,1,0>$ direction around these three Bragg positions (See Fig.~\ref{maps}). 
The reduction in the EDS intensity upon changing from the 
(0,0,2) to the (1,1,2) point is a factor of 
10 and the intensity around the (1,~1,~1) peak in turn is much weaker. As the polarization factor 
$\lvert\mathbf{Q\cdot u}\rvert^2$ stays nearly the same for the (0,0,2) to the (1,1,2) positions (see Table~\ref{tableq}), 
the difference in the EDS intensity is likely due to complex pattern of the ionic displacements.

\vspace{3 mm}
\begin{table}[h]
\label{tableq}
\centering
\begin{tabular}{|c|c|c|c|c|c|}
\hline
$\mathbf{\tau}$ & $ \lvert\mathbf{\tau}\rvert $ ,\AA$^{-1}$ & $\angle(\mathbf{\tau}, <0,0,1>), {^ \circ }$  & $\lvert\mathbf{Q\cdot u}\rvert^2$  \\
\hline
001 &  1.46   & 0   & 2.1  \\
\hline
002 &   2.92   & 0   & 8.5  \\
\hline
112 &   3.00   & 14  & 8.4  \\
\hline
111 &   1.63   & 26  & 2.1  \\
\hline
330 &  2.15   & 90  & 0     \\
\hline
\end{tabular}
\caption{Auxiliary information on the geometry of our experiments on SBN-61:Ce. $\mathbf{\tau}$ stands for a 
         particular reciprocal point vector. 
         $ \lvert\mathbf{\tau}\rvert $ is a distance from the origin of reciprocal space to the 
         appropriate Bragg peak. $\lvert\mathbf{Q\cdot u}\rvert^2$ 
         is a square of the absolute value of the polarization factor. For simplicity, vector 
         $\mathbf{u}$ is normalized to unity. Mutual arrangements of the 
         vectors are sketched in Fig.~\ref{scatplane}.} 
\end{table}

At T~=~300~K, in the ferroelectric phase, the EDS becomes much more condensed 
underneath the Bragg peaks. This is easily observed in the data collected near the (0,~0,~2) Bragg peak (see Figs.~\ref{maps}a,b). 
The nearly 
temperature-independent halo around the (0,~0,~2) position is very weak as compared to the EDS. This scattering is likely due 
to contribution from acoustic phonons 
to the elastic intensity. Importantly, no additional asymmetric  in wavevectors component in diffuse scattering have been detected 
in the vicinity of the reciprocal lattice positions. This means that possible Huang scattering in SBN-61:Ce is negligible. 
As the maps of the elastic neutron scattering were collected around the Bragg peaks with different parities in (H,K,L), 
we conclude that 
rather high doping level by Ce did not cause considerable excessive strain in the crystal.
Thus the the temperature-dependent EDS originates from chain-like ferroelectric correlations. The behavior of this EDS in 
paraelectric (Section~\ref{para}) and ferroelectric (Section~\ref{ferro}) phases is considered below.

\subsection{Critical scattering in paraelectric phase}
\label{para}
\begin{figure*}
\includegraphics[width=0.9\columnwidth]{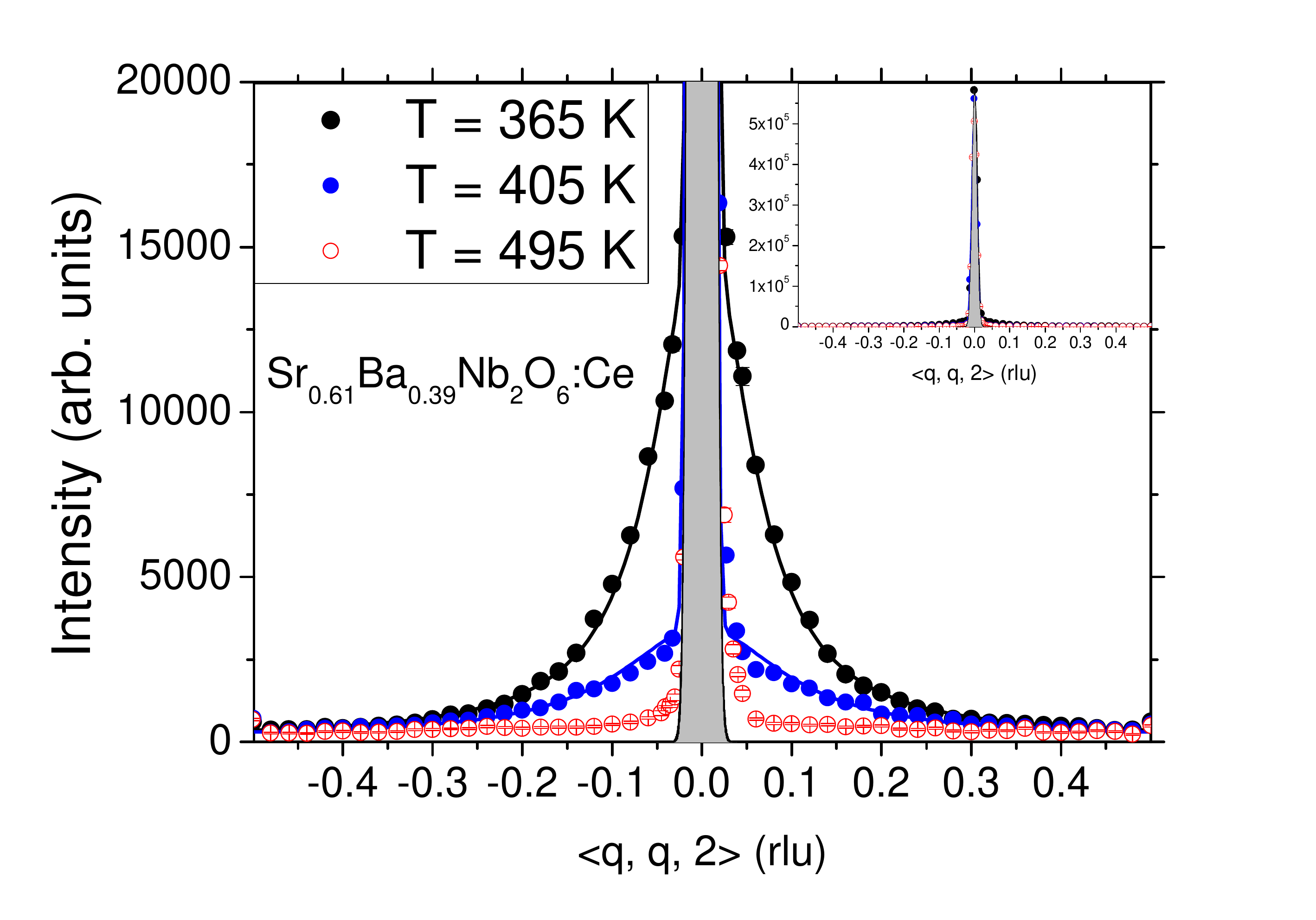}
\caption{Evolution of the elastic scattering in SBN-61:Ce taken along the $<$1, 1, 0$>$ direction above T$_\textrm{c}$. The 
         lines are the fits as described in the text. 
         The shaded areas emphasize the Bragg peaks. Their widths are defined by spectrometer resolution and 
         were calculated with ResLib~4.2~\cite{reslib}. The 
         data presented in Fig.~\ref{dsabove} is obtained 
         with open-20$'$-20$'$-20$'$ collimation. The inset shows the overall intensity in the scan.}
  \label{dsabove}
\end{figure*}

\noindent The elastic neutron scattering was studied as a function of temperature by performing scans along the 
$<$q, q, 0$>$ direction across the (0, 0, 1) and (0, 0, 2) 
Bragg peaks. This choice was dictated by the following reasons. Spontaneous polarization in SBN:Ce appears along 
the $c$-axis, thus the geometric condition for a transverse geometry is best 
fulfilled for the $<$q, q, 0$>$ direction with respect to the (0,~0,~L)-type Bragg peaks. Also, the neutron 
structure factors for the (0, 0, 1) and (0, 0, 2) Bragg peaks 
differ strongly. Probing simultaneously the scattering around such two reflections ensures reliable data 
collection and parametrization. 

Figure~\ref{dsabove} shows elastic neutron scans taken at several temperatures above the ferroelectric phase transition. 
These scans consist of two obvious components. The intense 
Bragg peak whose width is determined by experimental resolution and less intense but extended in wavevectors diffuse 
scattering centered at $q=0$. At higher temperatures this 
scattering is very broad and identifying its lineshape is difficult. On cooling towards T$_\textrm{c}$ the diffuse 
intensity rapidly increases. This tendency is just opposite to 
the behavior of IDS discussed in section~\ref{lowe}. Thus, the broader component of transverse scans shown in 
Fig.~\ref{dsabove} is elastic diffuse scattering, the EDS. 
 
The energy width of the EDS in SBN-61:Ce could not be resolved in our experiments. This requires an assumption on 
its intrinsic nature, whether this diffuse scattering is caused by 
static or dynamic fluctuations. Following the approach used for pure SBN-61~\cite{sbn61}, we assume the EDS is a 
critical scattering caused by dynamic, 
although very slow fluctuations. In this case, the neutron intensity can be approximated as 
\begin{equation}
\label{sf_eds}
I_{EDS}\sim\chi(q=0,T)\cdot T \cdot \frac{\kappa}{\kappa^2+q^2}, 
\end{equation} 

\noindent where $\chi(q=0,T)$ is the susceptibility associated with the EDS taken at the $q=0$, $\kappa=1/\xi$ is 
the inverse of the correlation length, and T is the temperature. 
In the range from 405~K down to $\sim350$~K the elastic scans collected around the (0,~0,~1) and the (0,~0,~2) 
positions were fitted to a Gaussian Bragg peak together 
with a Lorentzian curve convoluted with the Gaussian resolution width. Examples of such fits are presented 
in Fig.~\ref{dsabove} for T~=~405~K and T~=~365~K. 
Within this approach the susceptibility and the width of the EDS were obtained as a function of temperature. 

\begin{figure}
\includegraphics[width=0.5\columnwidth]{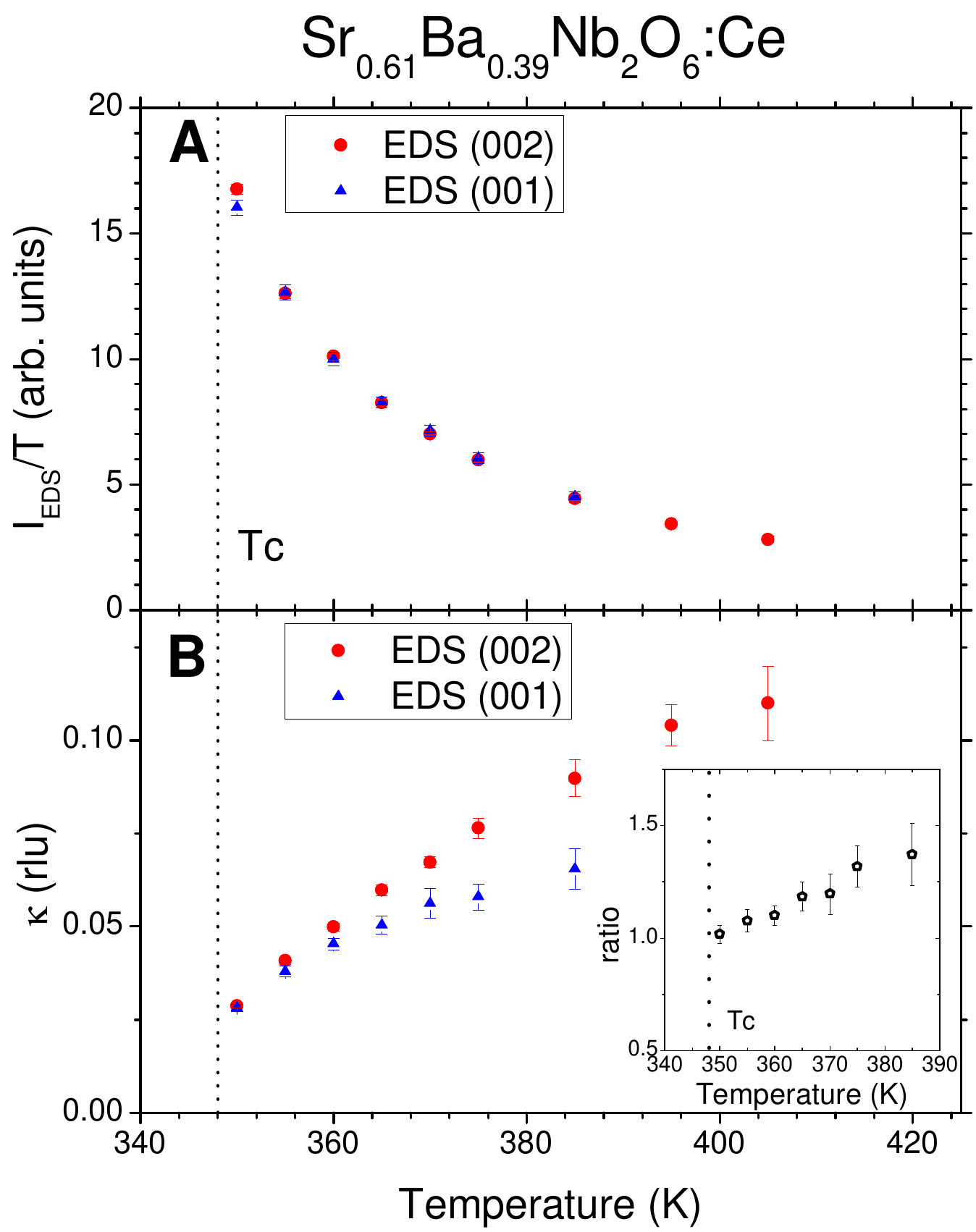}
\caption{a) Temperature dependence of the susceptibility associated with the EDS taken around the (0,~0,~1) and the 
           (0,~0,~2) Bragg peaks. The data collected around the 
           (0,~0,~1) position are multiplied by a single scale factor 15.8. 
         b) The half-width at half maximum $\kappa$ of neutron EDS from SBN-61:Ce measured in the vicinity of 
            the same reciprocal lattice positions. The inset shows the ratio of 
             the widths inferred from the data taken in the vicinity of the (0,~0,~2) and the (0,~0,~1) peaks. 
            Vertical dashed lines in both panels denote the ferroelectric transition.}
  \label{edspar}
\end{figure}

Figure~\ref{edspar}a shows the temperature dependence of the integrated intensity of the EDS scattering 
$I_{EDS}/T\sim\chi(q=0,T)$ from SBN-61:Ce measured around the (0,~0,~1) and the (0,~0,~2) positions. 
$\chi(q=0,T)$ probed around both Bragg reflections increases on approaching the phase transition from above. 
Furthermore, as discussed in Sections~\ref{lowe}\&\ref{shape}, 
we did not observe any other strongly 
temperature-dependent scattering in the vicinity of T$_\textrm{c}$. Taken together these observations imply 
that fluctuations producing the EDS 
account for the dielectric anomaly associated with ferroelectric transition of SBN-61:Ce. 

On approaching the phase transition one expects not only diverging susceptibility of critical fluctuations, but 
also condensing the associated intensity at the 
propagation vector. The width of the EDS from SBN-61:Ce, however, does not decrease to zero near T$_\textrm{c}$ 
as shown in Fig.~\ref{edspar}b. In contrast to the susceptibility, 
the width of the EDS probed in the vicinity of the (0,~0,~1) and the (0,~0,~2) Bragg peaks exhibits dissimilar 
temperature evolution. At higher temperature their ratio is about 1.3, 
but progressively becomes unity upon approaching T$_\textrm{c}$. This unusual behavior is possibly related with 
a complicated pattern of ionic displacements producing 
neutron EDS from SBN-61:Ce.

\subsection{Diffuse scattering in the ferroelectric phase: Domain Walls}
\label{ferro}
\begin{figure}
\includegraphics[width=0.5 \columnwidth]{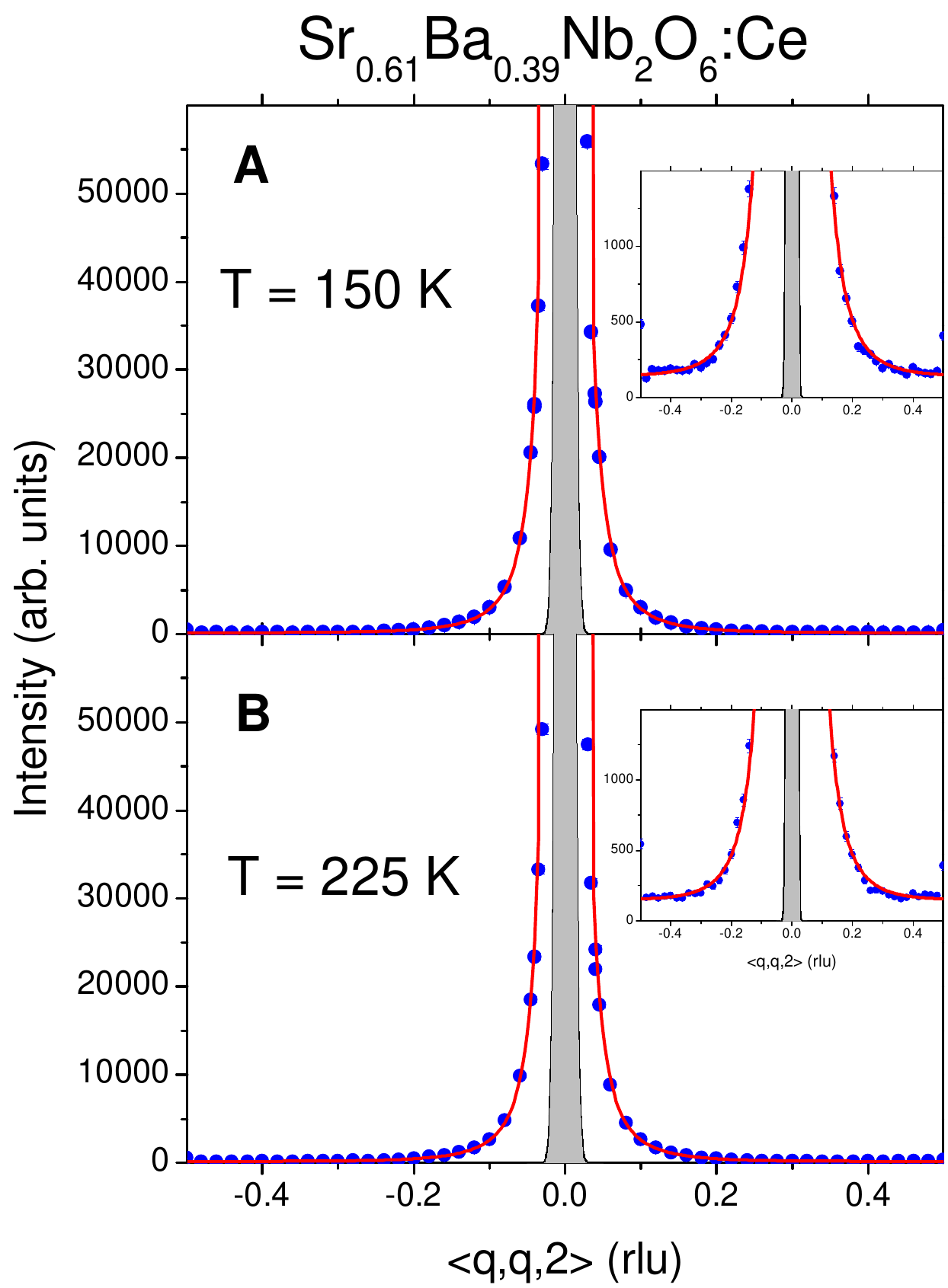}
\caption{The EDS below the phase transition in SBN-61:Ce shown together with fits to the domain wall model. 
         The data is shown by blue circles 
         and the fits to Eq.\ref{dwfinal} are given by red lines. The insets emphasize the low-intensity parts of the scans. 
         The shaded areas denote the 
         portion of the scans where the intensity would be influenced by the Bragg peak. Such data points 
         were not considered in the fits. 
         The data was taken with collimation as open-20$'$-20$'$-20$'$.}
  \label{dwshape}
\end{figure}

\noindent The distribution of neutron elastic scattering in the ferroelectric phase of SBN-61:Ce is very similar to that in 
the  paraelectric phase (see Section~\ref{shape}). However, below T$_\textrm{c}$ the EDS becomes considerably sharper 
in wavevectors and much more intense near $q=0$. Thus separating the EDS from Bragg peaks in the ferroelectric phase 
is more complicated. Furthermore, just below T$_\textrm{c}$ there are two contributions to the EDS. First, below T$_\textrm{c}$ 
the sample breaks up into ferroelectric domains. This causes intense diffuse scattering from the domain walls. 
In addition, just below T$_\textrm{c}$ there should be a contribution to the EDS caused by the critical scattering. 
The integrated susceptibility of the DS follows real part of dielectric permittivity $\varepsilon '$ closely for T$>$T$_\textrm{c}$. 
A similar behavior is expected for T$<$T$_\textrm{c}$. An inspection of Fig.~\ref{edspar}a suggests that the contribution from 
the critical scattering should become negligible for T$\le$300~K. We can expect the EDS from SBN-61:Ce is 
dominated by scattering from the ferroelectric domain walls below this temperature. With these assumptions 
we concentrate on the treatment of the EDS in the temperature range 150 - 300~K.

\begin{figure}
\includegraphics[width=0.5 \columnwidth]{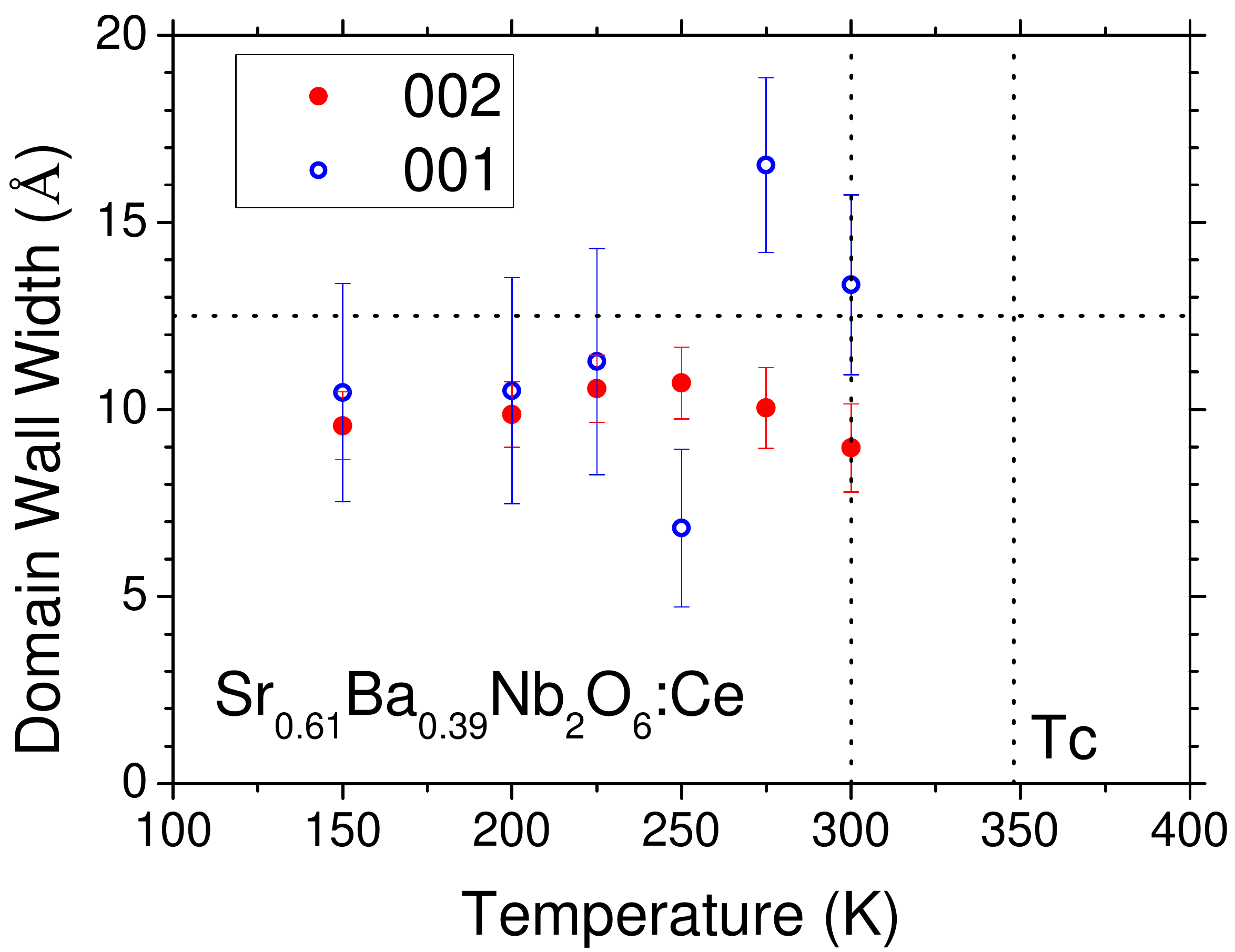}
\caption{The width of the ferroelectric domain walls in SBN-61:Ce in the temperature range 150~K-300~K. 
         The vertical lines denote the phase transition T$_\textrm{c}$ and the temperature where we assume 
         the contribution from critical scattering to the EDS becomes 
         negligible. The horizontal line denotes the value of the lattice constant.} 
  \label{dwthick}
\end{figure}

To analyze the lineshape of the EDS we need to recall the results on the diffuse scattering from the domain walls
~\cite{PhysRevLett36806,bruce_kdp,andrews_kdp,PhysRevB193630,sbn61}. For uniaxial ferroelectrics an analytical expression for 
the scattering intensity was obtained under the following assumptions. If no electric field is applied to a specimen, 
below T$_\textrm{c}$ the crystal breaks into the ferroelectric domains. They are expected to be macroscopically large 
along the direction of spontaneous polarization. These domains are likely to form the so-called 180$\rm^o$ 
structure with their normals orthogonal to the unique ferroelectric axis $c$. It is assumed that the spontaneous 
polarization is proportional to the ionic displacement field along $c$ as $P_x \sim u(x)$ where $u(x)=u_0\tanh(x/\lambda)$. This 
leads to the one-dimensional scattering function $S(q_x)=\frac{D\lambda^2}{\sinh^2(q_x\lambda/2)}$. Extending this result to 
three dimensions in wavevectors space requires further approximations. Namely, the domain walls are assumed to be randomly 
arranged in the a-b plane of the crystal~\cite{comment} and so interference effects are negligible in the scattering. 
Taken together, these arguments lead to the following scattering function 
for the diffuse scattering induced by the domain walls:
\begin{equation}
\label{dwfinal}
S(\mathbf{Q},\omega)=D\frac{1}{|q|}\frac{\lambda^2}{\sinh^2(q\lambda/2)}\frac{\kappa_z}{q^2+\kappa_z^2}\cdot\delta(\omega)  
\end{equation} 
  
\noindent here $\lambda$ is a half of the average thickness of the domain walls, $D$ is a constant proportional to their 
density, the reduced wavevector $q$ runs from respective Bragg peak, $\mathbf{q}=\mathbf{Q} \pm \mathbf{\tau}$. 
$\delta(\omega)$ accounts for elastic nature of the scattering. Lorentzian with small $\kappa_z$ accounts for large size of the domains along the unique ferroelectric direction. In our fits $\kappa_z$ was fixed 
to the value 0.0005 rlu to balance the calculation time and numeric precision of the modeling. 

The scattering function given by Eq.~\ref{dwfinal} was convoluted with the spectrometer resolution function by using 
ResLib4.2 library~\cite{reslib} and a constant 
background was added. The best fit results for SBN-61:Ce  are shown in Fig.~\ref{dwshape}.  
The model Eq.~\ref{dwfinal} reproduces the data taken around the Bragg peaks (0,0,1) and (0,0,2) in the temperature range 
220~K -- 300~K. The fitted curves describe the data spanning over more than 3 orders of magnitude in intensity. 
Such an agreement further validates 
our approach. The domain wall density (which is just the scale factor in the model) was found to be temperature independent. 
The temperature dependence of the width of the ferroelectric domain walls is shown in Fig.~\ref{dwthick}. In the addressed temperature 
range the width probed by neutron scattering near the (0,0,1) and (0,0,2) Bragg peaks is the same. This suggests that the pattern of 
displacements remains unchanged through the domain walls. The width of the domain walls in SBN-61:Ce is close to the lattice constant 
and is similar to the values found in other uniaxial 
ferroelectrics~\cite{PSSB:PSSB2220870121,PhysRevLett36806,PhysRevB193645,philmaga45911}. 

\section{Discussion and Conclusions}

\noindent As a rule, phase transitions are related with a condensation of low-energy modes. In particular, on approaching 
the ferroelectric phase transition the transverse polarization fluctuations should slow down and the associated 
susceptibility diverges. The spontaneous polarization appears in SBN crystals along the $c$- axis. 
Thus, in pure and Ce-doped SBN-61~\cite{sbn61} we studied in details the low-energy transverse modes propagating along 
the $<$q, q, 0$>$ direction. Getting a few steps forward, the scattering properties of the two crystals are very similar. 

We were able to model the intensity of the TA phonon in Ce-doped SBN-61 assuming DHO lineshape and sinusoidal 
dispersion relation (see \ref{lowe}). This approach 
allows for more reliable estimates of the damping, but is different from previously employed in other 
studies of SBN~\cite{sbn61,PSSB:PSSB2220870121}. 
To compare more closely the TA dispersion in SBN-61:Ce with that in 
SBN crystals of other compositions the phonon frequencies were also 
obtained via simple Lorentzian fits. These results are shown in Fig.~\ref{parinel}D for several temperatures. 

The critical scattering observed in the paraelectric phase of Ce-doped SBN-61 is consistent with chain-like 
correlations of ionic displacements. However, the displacements pattern has additional modulation leading to 
unexpected difference in width of the EDS probed around different Bragg positions but along the same direction 
in reciprocal space. Upon approaching the transition temperature this width of the EDS decreases and 
the associated susceptibility increases. 
   
In the ferroelectric phase the EDS from Ce-doped SBN-61 is produced by the domain walls. The width of this DW 
inferred from the data taken around the (0,0,1) and (0,0,2) Bragg reflections is essentially the same and 
well below T$_\textrm{c}$ does 
not depend noticeably on the temperature. 

In summary we compare the properties of unmodified and Ce-doped SBN-61 as seen by neutron scattering.
The dispersion curve of the TA phonon propagating along the $<$q, q, 0$>$ direction does not change upon 
doping by Ce~\cite{sbn61,PSSB:PSSB2220870121}. The critical scattering observed in the paraelectric phase 
of Ce-doped SBN-61 is similar to that one from undoped SBN 
crystals and in both cases its distribution is consistent with chain-like correlations of ionic 
displacements. An additional charge disorder introduced by Ce$^{3+}$ ions does 
not lead to noticeable change in ferroelectric DW thickness which was found to be $\sim10$~\AA~ for both 
crystals.  Despite the significant differences in macroscopic properties, the lattice dynamics, 
critical dynamics and domain structure of Ce-doped SBN-61 surprisingly is almost identical to that of SBN-61.

\section{Acknowledgments}
The experiments were performed at the SINQ facility at PSI. 
We thank Markus Zolliker and Walter Latscha for excellent technical support.

\bibliographystyle{apsrev4-1}
\bibliography{sbn_ce}{}

\begin{thebibliography}{38}
\bibitem [{com()}]{comment}%
  \BibitemOpen
  \href@noop {} {}\bibinfo {note} {The assumptions on the shape of the
  ferroelectric domains in SBN are in agreement with observations made by
  piezoresponse force microscopy~[38]}\BibitemShut {NoStop}%
\end{thebibliography}
\end{document}